# Bending and wrinkling as competing relaxation pathways for strained free-hanging films

P. Cendula, [1,3] S. Kiravittaya, [2] Y. Mei, [1] Ch. Deneke[1] and O. G. Schmidt[1]

[1] *Institute for Integrative Nanosciences, IFW Dresden, Helmholtzstr. 20, D-01069 Dresden, Germany*
[2] *Max-Planck-Institut für Festkörperforschung, Heisenbergstrasse 1, D-70569 Stuttgart, Germany*
[3] *DCMP, Faculty of Mathematics and Physics, Charles University in Prague, Czech Republic*

An equilibrium phase diagram for the shape of compressively strained free-hanging films is developed by total strain energy minimization. For small strain gradients $\Delta\varepsilon$, the film wrinkles, while for sufficiently large $\Delta\varepsilon$, a phase transition from wrinkling to bending occurs. We consider competing relaxation mechanisms for free-hanging films, which have rolled up into tube structures, and we provide an upper limit for the maximum achievable number of tube rotations.

A thin solid film, subject to compressive strain, can either bend [1-2] or wrinkle [3-5], if fixed at one end and free otherwise. Whether the film bends or wrinkles sensitively depends on the built-in strain gradient across the film thickness. Intuitively speaking, if the strain gradient is large, the film bends into a curved structure, whereas for a small or zero strain gradient the film forms wrinkles. Interestingly, the competing mechanisms of wrinkling and bending, as the strain gradient inside the film changes have not been quantified so far. This circumstance is even more surprising since a variety of fundamental investigations as well as applications based on bent [1-2,6] and wrinkled [3-5,7-8] films have been put forward. The roll-up of a strained film into a cylindrical geometry seems particularly appealing [9-10], since size, orientation and number of rotations of a micro- or nanotube become well-controlled and predictable entities. These virtues have led to exciting perspectives both in fundamental research [11-15] and also with respect to applications [16-19].

In this Letter, we perform an energetic comparison between bent/rolled and wrinkled films, and we generate a quantitative *a priori* phase diagram for the formation of bent and wrinkled structures. Based on these two competing strain relaxation pathways, we are able to provide an upper fundamental limit for the number of film rotations as the free-hanging film progressively increases in length.

Figure 1(a) shows a schematic illustration of a partially released bilayer film, consisting of two layers with thicknesses $d_1$ and $d_2$, which are subject to biaxial strain $\varepsilon_1$ and $\varepsilon_2$, respectively. The film is free-hanging over a distance $h$, and is in an unrelaxed strained state over the whole length $L$. Experimentally, the free-hanging film can be fabricated by selectively etching away a sacrificial layer (as indicated in the figure) [3,9-10], but other procedures to create free hanging films are also possible [4]. The released portion of the film is free to elastically relax, constrained only by the fixed boundary where the film attaches to the substrate/sacrificial layer. The unreleased part of the film is still firmly bonded to the substrate/sacrificial layer. The layers are assumed to have equal isotropic linear elastic properties with Young's modulus $Y$ and Poisson's ratio $\nu$. The average strain and strain gradient of the bilayer are defined as $\bar{\varepsilon} = (\varepsilon_1 d_1 + \varepsilon_2 d_2)/(d_1 + d_2)$ and $\Delta\varepsilon = \varepsilon_2 - \varepsilon_1$ respectively. The initial elastic energy (given per unit area) of the film is $E_0 = Y\left(d_1\varepsilon_1^2 + d_2\varepsilon_2^2\right)/(1-\nu)$.

We consider the initial stage of strain relaxation for the case $\Delta\varepsilon > 0$ in Fig. 1(b). The film bends and adapts a uniform inner radius $R$. The energy calculation is performed in a cylindrical coordinate system $(x, t, r)$ with the origin at the outer surface of the bent film in Fig. 1(b). We adopt the approach from Ref. [20] to estimate the equilibrium radius and elastic energy of a film subject to a certain strain gradient. Since the film is still firmly attached to the sacrificial layer, and we assume $L \gg R$, there is no relaxation in the $x$ direction [21] (plane strain condition in $x$). Therefore $\varepsilon_{xi} = \varepsilon_i$, $i = 1,2$ for layer no. 1 and no. 2, respectively.

In the tangential direction, the strain can relax by bending and the final tangential strain can be written as $\varepsilon_{ti} = \varepsilon_i + c - (r - r_b)/R$, where $c$ is a uniform strain and $r_b$ indicates the location of the neutral plane, where the bending strain component is equal to zero. Since the layers are thin, the stress through their thickness in the radial direction has to be zero ($\sigma_r = 0$) at equilibrium [22], implying by Hooke's law for the strain in the radial direction $\varepsilon_{ri} = -\nu(\varepsilon_{ti} + \varepsilon_{xi})/(1-\nu)$. From the condition of zero total bending force on the film we derive $r_b = d_2(1+\delta)/2$, with $\delta = d_1/d_2$.

The total elastic energy of the bent film $E_{bent}$ is calculated by integrating the elastic strain energy density from the outer to the inner film surface. By minimizing the energy with respect to the remaining unknown parameters $c$ and $R$, we obtain the equilibrium uniform strain $c_{eq} = -\eta(\delta\varepsilon_1 + \varepsilon_2)/(1+\delta)$ and the equilibrium tube radius $R_{eq} = \rho d_2/(6\eta\Delta\varepsilon\delta)$, where $\rho = (1+\delta)^3$ and $\eta = 1+\nu$. Subsequently, $R_{eq}$ and $c_{eq}$ are used to calculate the equilibrium minimum elastic energy of the rolled structure $E_{bent}(R_{eq}, c_{eq})$. This value is normalized to $E_0$ and will be later compared with the wrinkle energy.

Throughout the article, we consider a typical bilayer consisting of 10 nm $In_{0.1}Ga_{0.9}As$ with $\varepsilon_1 = -0.71\%$ and 10 nm GaAs with $\varepsilon_2 = 0\%$, and equal $Y = 80$ GPa, $\nu = 0.31$. For this case, the equilibrium radius and minimum energy are $R_{eq} \approx 1.4$ μm and $0.43E_0$, respectively.

For the calculation of the elastic energy of the wrinkled structure, we extend a previous formulation of a single layer [5] to the bilayer film. We parameterize the vertical deflection of the wrinkle as $\zeta(x,y) = Af(y)\cos(kx)$, where $A$ is the maximum amplitude of the wrinkle at the free end [see Fig. 1(c)], $k = 2\pi/\lambda$ is the wrinkle wavenumber in the $x$ direction ($\lambda$ is wavelength of the wrinkle) and $f(y) = [1-\cos(\pi y/h)]/2$. The fixed boundary forces the left side of the wrinkle to be clamped, i.e. $\zeta(x,0) = 0$, $\zeta_{,y}(x,0) = 0$, where the partial derivative is denoted by a comma, and this is satisfied naturally by our choice of $\zeta$. The strain $\varepsilon_{\alpha\beta}^{(i)}$ and stress $\sigma_{\alpha\beta}^{(i)}$ are defined according to the large deflection (Föppl–von Kármán) plate theory [22], where $\alpha, \beta = x, y$. The in-plane displacement $u_x$ is approximated by modifying the result of the in-plane equilibrium for our shape [23], $u_x = (k/8)[Af(y)]^2 \sin(2kx)$. The film is free to move in $y$, so we take $u_y = \gamma y$ (neglecting an $x$ dependence) with parameter $\gamma$, denoting the magnitude of relaxation in the $y$ direction. The total wrinkle elastic energy $E_w$ can be decoupled [22] into a stretching energy $E_S$ and a bending energy $E_B$ (modified for the bilayer system [24]). The wrinkle energy $E_w$ is averaged over one wavelength, $L = \lambda$, and is numerically minimized with respect to $A$, $\lambda$ and $\gamma$. This minimum of the total energy is equivalent to the mechanical equilibrium within Föppl–von Kármán theory [22]. The interplay between the stretching and bending energies determines the equilibrium wrinkle periodicity and amplitude.

The wrinkle energy as a function of wrinkle length for our structure is given in Fig. 1(d). Below a critical length [25], referred to as the "critical wrinkle length" $h_{cw} \approx 2.57 d_2/\sqrt{-\bar{\varepsilon}}$ [26], energy minimization provides only a trivial minimum [4,27] of the wrinkle energy with $A = 0$ and $\lambda \to \infty$. This corresponds to the "planar" relaxation in the $y$ direction only [dashed line in Fig. 1(d)]. For our typical structure, the obtained value of $h_{cw}$ is about 450 nm and the planar energy $0.68 E_0$. Beyond $h_{cw}$, wrinkling can occur with nonzero amplitude and energy lower than the planar value. For very large wrinkles, the normalized wrinkle energy $E_w/E_0$ reaches an asymptotic value $0.60 E_0$. The amplitude and wavelength of wrinkles are plotted in Fig. 1(e) and scale as $\sim h^{0.65\pm0.05}$ in the range $h = (1-100)$ μm, similar to the previously reported scaling for the same structure [5] and slightly different from the $h^{0.5}$ reported for the general wrinkling phenomena [28].

The preferable shape of the free-hanging film of length $h$ is determined by comparing the normalized energy of the bent $E_{bent}(R_{eq}, c_{eq})$ and wrinkle $E_w(A, \lambda, \gamma)$ shapes. For our typical structure, the energy of the wrinkle [$0.60E_0$ to $0.68E_0$, see Fig. 1(e)] is always larger than the energy of the bent structure ($0.43E_0$).

To extend our considerations, we systematically change $\varepsilon_1$ and $\varepsilon_2$ and calculate the favourable shapes as a function of $h$ and strain gradient, as shown in the phase diagram, Fig. 2. The strain gradient $\Delta\varepsilon$ and etching depth $h$ are varied, while we fix the average strain to the value of our typical structure ($\bar{\varepsilon} = -0.36\%$). The boundary between these two shapes is shown as a solid line in Fig.2. For example, for $\Delta\varepsilon = 0.20\%$ and $\bar{\varepsilon} = -0.36\%$, bending will be favored only until $h$ is increased to $\approx 700$ nm. Beyond this length, the wrinkle becomes favorable geometry as it acquires lower energy than the bent structure. If we consider higher average strain, for example $\bar{\varepsilon} = -1.0\%$ (dashed boundary in Fig. 2), the phase boundary curve moves upward and the wrinkling region is enlarged.

We can use our model to estimate the maximum number of rotations $N_{max}$ for a film released from the substrate by progressive underetching [9-10]. Consider the tube, which has already rolled up over some rotations with non-interacting windings. The tube radius linearly increases with the number of rotations $N$ (continuous variable) as $R_N = R_{eq} + N(d_1 + d_2)$. We consider an additional portion of the released layer of length $H$ as outlined in Fig. 3(a). The film has two pathways to relax. Either the film continues to roll up with radius $R_N$ [Fig. 3(b)] or forms wrinkles [Figs. 3(c) and (d)]. These two processes are energetically compared to calculate the maximum number of rotations.

When the film of length $H$ is rolled onto the tube with outer radius $R > R_{eq}$, the energy stored in the layer increases as the radius increases. For infinite radius the energy approaches the value of a planar relaxed film. The energy of a rolled up film as a function of its radius is given in the inset of Fig. 4.

For the energy of the wrinkled film, the previous formulation needs to be slightly modified. The right wrinkle/tube boundary (TB) has to be smoothly linked to the tube, i.e. $\zeta(x,H) = 0$ and $\zeta_{,y}(x,H) = 0$, see Fig. 3 (c). As before, the total wrinkle energy will be compared with the minimum elastic energy of the rolled up structure. For this wrinkle shape, the critical wrinkling length is doubled, $H_{cw} = 2h_{cw}$.

The maximum radius of the outer tube rotation $R_N = R_{max}$ is reached when the wrinkle energy of the free standing film becomes lower than the roll-up energy (see typical energy comparison in the inset of Fig. 4). For this comparison we take $H = 3H_{cw}$, since for this length the strain energy is largely relaxed compared to the planar value (see Fig. 1 (d) for wrinkle with half size domain $h = 3h_{cw}$).

The maximum number of rotations is determined by the relation $N_{max} = (R_{max} - R_{eq})/(d_1 + d_2)$. For our typical structure we obtain $N_{max} \approx 510$, point B in Fig. 4. For a broad range of strains, the maximum numbers of tube rotations are shown in Fig. 4. Rolled up films with less than one rotation are obtained for $\Delta\varepsilon < -\bar{\varepsilon}$, otherwise $N_{max}$ increases rapidly when the magnitude of average strain is decreased towards zero for non-zero $\Delta\varepsilon$.

We note that (1) altering our assumption on the wrinkle shape (e.g. from trigonometric to polynomial function) and varying elastic constants within realistic values do not qualitatively change our results. (2) Our theory is valid for asymmetric bilayers as well, but the phase diagram and the maximum number of rotations will be quantitatively modified. (3) For tensile average strain $\bar{\varepsilon} > 0$, no physical minimum of wrinkling energy exists for our model, consistent with observations of wrinkles only for compressive strains near the fixed boundary [3-5,28]. (4) In typical experiments the layer is partially released by selectively etching away a sacrificial buffer layer [3,9-10]. If the amplitude of the wrinkle becomes too large, the film may touch the substrate. In this case our model does not apply, because the film-substrate interaction energies might be larger than the elastic energy relaxation through bending and/or wrinkling. The same applies for $\Delta\varepsilon < 0$, where the film rolls downwards towards the substrate surface [29].

For the estimation of the maximum number of tube rotations, we assume non- or only weakly interacting windings. If the windings are tightly bonded together, the infinitesimally small incremental increase in $H$ during roll-up does not allow a sufficient length for wrinkling and thus the number of rotations is not limited within the framework of our model. Both cases, tightly bonded [12, 18] and non- or weakly interacting windings [30], have been reported in the literature. The number of rotations might be influenced by certain process parameters such as finite fluid flow during underetching. In this way the maximum number of rotations might be increased if the fluid flow is applied along the roll-up direction. For short films, i.e. $L \ll R$, wrinkling might not occur due to the full relaxation in this direction (plane stress) [20]. As a result, there is no limit of the number of rotations within the framework of our model. Considering $H$ much larger than $3H_{cw}$ (which approaches the saturation wrinkle energy) will lead to a decrease of $N_{max}$ on the order of 10%.

Systematic experimental data to explore the maximum number of tube rotations are missing. The maximum reported values $N_{max}$ for $In_{0.33}Ga_{0.67}As/GaAs$ are 30-40 rotations [7,31], about one order of magnitude below our prediction $N_{max} \approx 250$ for this system. This might be due to specific processing parameters present in experiments, for instance a misalignment of the rolling direction from the $y$ axis, an inhomogeneous etching front and loose rotations [7]. Therefore, the maximum experimentally obtained number of rotations have always been lower than our theoretical estimate.

The phase diagram in Fig. 2 and the estimations of the number of tube rotations in Fig.4 can be used as a predictive tool for the deliberate design of rolled up/wrinkled structures. Our theory is not restricted to any material and can be easily extended to multilayer systems.

In conclusion, we have performed an energetic comparison between the bending and wrinkling of compressively strained free-hanging films, and we have drawn the phase diagram for the preferential shape of the film as a function of length, average strain and strain gradient. We have applied our theory to estimate the maximum number of tube rotations during a roll-up process. We are aware of the limitations of our model, which we have carefully discussed and taken into account for all interpretations. Our considerations provide the theoretical framework to fundamentally understand bending and wrinkling of free-hanging films attached to one fixed boundary. Since such layers have gained substantial relevance for applications, our work is of practical interest for many materials and material combinations as well as for different geometries and length scales.

The authors acknowledge D. Bourne, B. Schmidt, F. Cavallo, R. Costescu, A. Rastelli, R. Songmuang, E. J. Smith, D. J. Thurmer and V. Holy for fruitful discussions. This work is financially supported by the BMBF (03N8711).

Email: p.cendula@ifw-dresden.de

**FIGURE CAPTIONS**

FIG. 1. (Color online) Schematics of (a) free-hanging bilayer film, (b) bent film with inner radius $R$ and (c) wrinkled structure with deflection profile $\zeta(x,y)$, amplitude $A$ and wavelength $\lambda$. (d) Wrinkle energy (solid line) and energy of planar relaxation (dashed line) as a function of $h$. (e) Wrinkle wavelength (solid line, left axis) and amplitude (dashed line, right axis) as function of wrinkle length $h$. Vertical dot-dashed line marks the critical wrinkle length $h_{cw}$.

FIG. 2. (Color online) Phase diagram of favorable film shapes based on the energetic comparison between bent and wrinkled structures. Solid curve shows the boundary between bent and wrinkled shapes for our typical structure. $R_{eq}$ is shown for the bent structure and wavelength $\lambda$ for the wrinkled structure. Dashed curve shows the phase boundary curve for $\bar{\varepsilon} = -1.0\%$.

FIG. 3. (Color online) Schematic illustration of the mechanism, which ceases the roll-up process. (a) A strip of length $H$ can (b) roll onto the outer part of the tube or (c) wrinkle, depending on the final energy of the system. (d) Assumed 3D wrinkle profile between fixed boundary (left) and tube boundary (TB, right).

FIG. 4. (Color online) Contour of the maximum number of tube rotations $N_{max}$ as a function of average strain and strain gradient for the considered wrinkle length $H = 3H_{cw}$. The inset shows a comparison between the strain energies of roll-up and wrinkling for $\bar{\varepsilon} = -1.0\%$ and $\Delta\varepsilon = 2.0\%$ (point A in the diagram). Point B denotes our typical strain combination as specified in the text. The dashed line denotes typical strain combinations for bilayers, where only one layer is compressively strained initially.

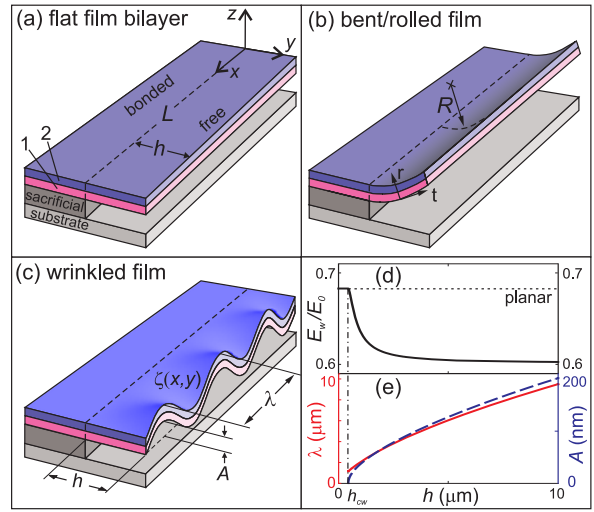

Fig.1

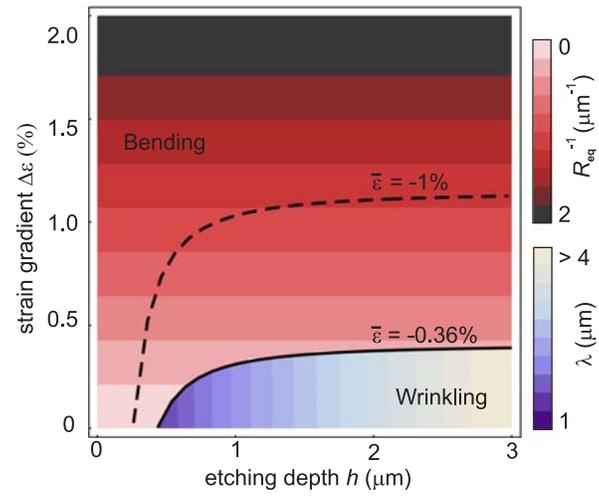

Fig.2

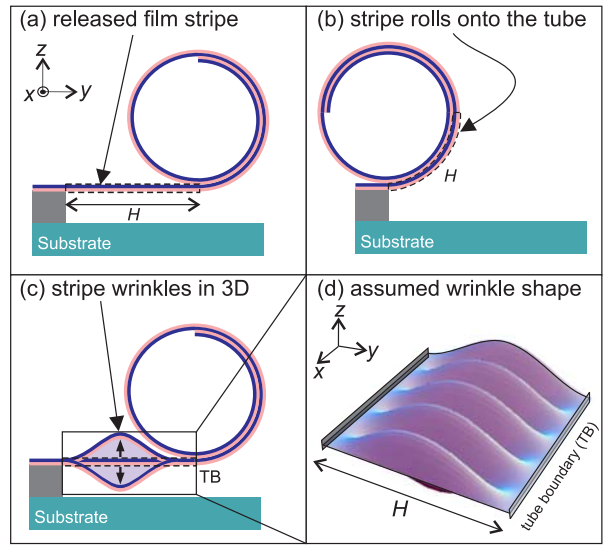

Fig.3

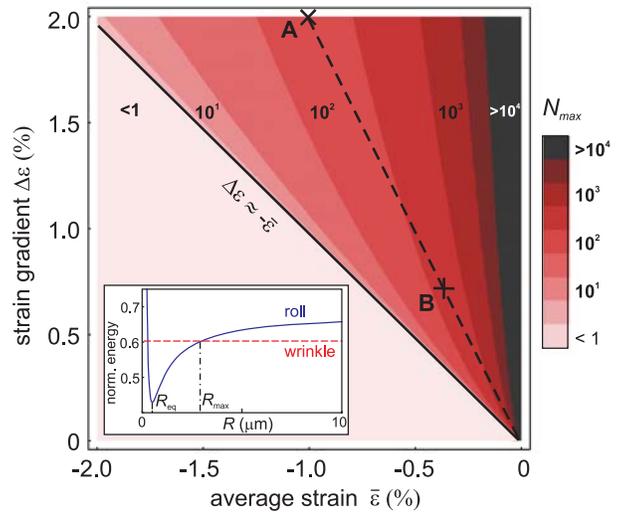

Fig.4